\documentclass[aps,prb,twocolumn,superscriptaddress,floatfix,longbibliography]{revtex4-2}

\usepackage{amsmath,amssymb} 
\usepackage{bm} 
\usepackage{graphicx} 
\usepackage{comment} 
\usepackage{textcomp} 

\usepackage{enumitem}
\setlist{noitemsep}

\usepackage{xcolor} 
\definecolor{lightgray}{gray}{0.6}
\definecolor{medgray}{gray}{0.4}

\usepackage{hyperref}
\hypersetup{
colorlinks=true,
urlcolor= blue,
citecolor=blue,
linkcolor= blue,
}

\newif\ifptitle
\newif\ifpnumber
\newcounter{para}

\ptitletrue  
\pnumbertrue  

\newcommand{\mytitle}{An assumption required to reproduce inclusive hadronic cross section data}

\begin{document}

\title{\mytitle}

\author{Scott Chapman}
\email[]{schapman@chapman.edu}
\affiliation{Institute for Quantum Studies, Chapman University, Orange, CA  92866, USA}

\date{\today}

\begin{abstract}
For 50 years, Standard Model calculations have successfully reproduced inclusive hadronic cross section data from high-energy $e^+e^-$ collisions.  However, these calculations have always assumed that there is no ``flavor interference'' between $e^+e^-\to u\bar{u}\to X$ and $e^+e^-\to d\bar{d}\to X$ events (where $X$ is the same hadronic state).  That assumption is questioned. 
\end{abstract}

\maketitle

Calculating an inclusive cross section involves taking the sum of exclusive cross sections over a complete set of orthogonal final states.  It has been argued that when calculating the inclusive cross section, any complete set of orthogonal final states can be used; they will all generate the same result.  The argument further posits that there is no requirement that a final state must be observable.  In that case, three of the possible definitions of a complete set of ``final'' states for $e^+e^-$ collisions are (a) an off-shell photon and a $Z$ boson, (b) all flavors of lepton-antilepton and quark-antiquark pairs, or (c) all flavors of lepton-antilepton pairs and all states involving hadrons.  According to the above argument, calculation of the inclusive cross section for $e^+e^-$ collisions generates the same result no matter which of the three sets is used. 

But this is not correct.  If set (a) is used, the Z boson and the photon are treated as orthogonal final states, so there is no interference between them.  If set (b) is used, then for $\sqrt{s}$ near the mass of the $Z$ boson, inteference between the $Z$ boson and the photon is included (see for example eq (1.34) of \cite{Z-summary}).  The inclusion of interference is justified by the fact that there exist lepton final states that could have been generated either by $e^+e^-\to Z \to l\bar{l}$ or by $e^+e^-\to \gamma^* \to l\bar{l}$ (interference also occurs for quark final states).  Due to this difference as to whether or not interference is included, calculation of the inclusive cross section for $e^+e^-$ collisions generates different results, depending on whether complete orthogonal set (a) or (b) is used.  

Despite this well known effect, it has been argued that there would be no difference in inclusive cross section calculations using complete set (b) vs. set (c).  Such an equality could only arise if there was no possibility of interference between different $q\bar{q}$ states in the set (c) calculation.  If set (c) included a final hadronic state $X$ that could have been created either by $e^+e^-\to q\bar{q}\to X$ or by $e^+e^-\to q'\bar{q}'\to X$, where $q\ne q'$, then ``flavor interference'' would need to be taken into account for that state when doing a set (c) calculation.  The argument that calculations using set (b) and set (c) generate the same result relies on there being no hadronic states $X$ like this in set (c).  

But that is clearly not the case.  The hadronic state $X=\pi^+\pi^-$ is one that can be created either by $e^+e^-\to u\bar{u}\to \pi^+\pi^-$ or by $e^+e^-\to d\bar{d}\to \pi^+\pi^-$.  More generally, any state comprised of two jets, where each jet has no net beauty, charm, strangeness or baryon number is equally likely to have been produced from a $u\bar{u}$ or a $d\bar{d}$ intermediate state.  Generalizing further, there is a very large class of hadronic states $X$ that can be produced via either a $u\bar{u}$ or a $d\bar{d}$ intermediate state.  When calculating the contribution of these states to the inclusive cross section using set (c), each diagram with an intermediate $u\bar{u}$ should be added to the corresponding diagram with an intermediate $d\bar{d}$ before squaring. After including inteference in this way, there is no reason to expect that inclusive cross section calculations using set (b) and set (c) should generate the same result.

Given the choice between calculations using sets (b) and (c), the set (c) calculation is more consistent with other quantum mechanical interference calculations, since all of the final states in the set are observable. The problem with using set (c) is a practical one.  Hadronization is a highly nonperturbative process, making calculation of the inclusive cross section impossible analytically.  Some new numerical algorithms would be needed to calculate the cross section in the presence of flavor interference.  Despite these calculational difficulties, inclusive hadronic cross section calculations using set (b) (the usual choice) should be viewed with skepticism.  

A different justification for using set (b) in calculations would be if there was reason to believe that the effects of flavor interference were insignificant.  To qualitatively address that, it is instructive to revisit the hadronic state $X=\pi^+\pi^-$.  As mentioned above, $X$ can be created either by $e^+e^-\to u\bar{u}\to \pi^+\pi^-$ or by $e^+e^-\to \bar{d}d\to \pi^+\pi^-$.  The matrix elements for these two processes should be added before being squared.  Even though hadronization is nonperturbative, isospin symmetry means that the $u\bar{u}\to \pi^+\pi^-$ and $\bar{d}d\to \pi^+\pi^-$ parts of the two matrix elements should be approximately equal. Therefore, the sum of matrix elements should have a partial cancellation due to the opposite signs of the electric (or Z boson) charges in the $e^+e^-\to u\bar{u}$ and $e^+e^-\to \bar{d}d$ parts of each diagram. For high CM energies below the Z pole (e.g. $20<\sqrt{s}<45$ GeV), comparing $(\tfrac{2}{3}-\tfrac{1}{3})^2$ to $(\tfrac{2}{3})^2+(\tfrac{1}{3})^2$, one can see that the flavor interference of this sum is significant and destructive. Repeating the analysis for other hadronic states $X$ with increasing complexity, significant interference is seen, sometimes destructive, sometimes constructive.

The situation is further complicated by the fact that interference can also be generated from $s\bar{s}$ intermediate states.  That means that the net interference effect may be different for low CM energies (e.g. below 3 GeV) than for high energies (e.g. above 10 GeV).  The reason is because fewer $s\bar{s}$ pairs (relative to $u\bar{u}$ and $d\bar{d}$ pairs) are created for the former than for the latter. 

A different justification for using set (b) rather than set (c) would be a proof that the large interference effects experienced by set (c) states all cancel.  Such a proof does not appear to exist in the literature.

What about interference involving the heavier quark flavors?  Consider the exclusive process $e^+e^- \to B^+B^-$.  The final hadronic state $X=B^+B^-$ could be generated by either $e^+e^- \to \bar{b}b \to B^+B^-$ or by $e^+e^- \to u\bar{u} \to B^+B^-$. But the latter would require nonperturbative generation of a $\bar{b}b$ pair.  Such pair creation should have a very small contribution since $2m_b\gg \Lambda_{\rm QCD}$.  On the other hand, since $2m_u\ll \Lambda_{\rm QCD}$, a $u\bar{u}$ pair can easily be generated, so $e^+e^- \to B^+B^-$ should be generated almost exclusively from the $e^+e^- \to \bar{b}b$ subprocess.  That same argument would hold for any hadronic final state in which at least one jet carries a beauty quantum number.  This suggests that the usual approach of ignoring flavor interference should be a good approximation for hadronic states involving beauty or charm quarks, since the masses of those quarks are much larger than $\Lambda_{\rm QCD}$.

In summary, past calculations of inclusive hadronic cross sections have implicitly assumed that there is no interference (or negligible interference) between $e^+e^-\to u\bar{u}\to X$, $e^+e^-\to d\bar{d}\to X$ and $e^+e^-\to s\bar{s}\to X$ events, where $X$ is the same hadronic state.  That assumption is questioned.

If this kind of ``flavor interference'' does need to be taken into account, it is possible that a correct calculation of the inclusive hadronic cross section would not agree with experimental data.  That could then imply that something beyond the Standard Model would be needed to reproduce experimental data.

To settle this question, a calculation that fully takes into account u/d/s flavor interference should be performed.

\end{document}